\documentclass{osa-article}

\journal{oe}
\articletype{Research Article}
\usepackage{lineno}
\usepackage{multicol}
\usepackage{multirow}
\begin{document}

\title{Single-pixel imaging of a translational object}

\author{Shijian Li,\authormark{1,$\dagger$} Yan Cai,\authormark{2,$\dagger$} Yeliang Wang,\authormark{1} Xu-ri Yao,\authormark{2,4} and Qing Zhao \authormark{2,3,5}}

\address{\authormark{1}School of Integrated Circuits and Electronics, Beijing Institute of Technology, Beijing 100081, China\\
\authormark{2}Center for Quantum Technology Research and Key Laboratory of Advanced Optoelectronic Quantum Architecture and Measurements (MOE), School of Physics, Beijing Institute of Technology, Beijing 100081, China \\
\authormark{3}Beijing Academy of Quantum Information Sciences, Beijing 100193, China \\
}
\email{\authormark{4}yaoxuri@bit.edu.cn\\
\authormark{5}qzhaoyuping@bit.edu.cn} %% email address is required
\address{\authormark{$\dagger$}These two authors contribute equally to this work \\
}


\begin{abstract}
Image-free tracking methods based on single-pixel detectors (SPDs) can track a moving object at a very high frame rate, but they rarely can achieve simultaneous imaging of such an object. In this study, we propose a method for simultaneously obtaining the relative displacements and images of a translational object. Four binary Fourier patterns and two differential Hadamard patterns are used to modulate one frame of the object and then modulated light signals are obtained by SPD. The relative displacements and image of the moving object can be gradually obtained along with the detection. The proposed method does not require any prior knowledge of the object and its motion. The method has been verified by simulations and experiments, achieving a frame rate of 3332~$\mathrm{Hz}$ to acquire relative displacements of a translational object at a spatial resolution of 128 × 128 pixels using a 20000-$\mathrm{Hz}$ digital micro-mirror device. This proposed method can broaden the application of image-free tracking methods and obtain spatial information about moving objects.
\end{abstract}

%%%%%%%%%%%%%%%%%%%%%%%%%% body %%%%%%%%%%%%%%%%%%%%%%%%%%
\section{Introduction} 
Tracking and imaging fast-moving objects have significant application prospects in navigation, biomedical, computer vision, and other fields. The two main reasons for the deterioration of imaging quality for a fast-moving object are blurring caused by motion and the low signal-to-noise ratio caused by the high frame-rate shooting. The high-speed camera\cite{kondo2012development} was invented to capture moving objects with a very high frame rate and relatively high signal-to-noise ratio, but it is expensive, and the data flux is very high. Various tracking and imaging methods for moving objects based on spatial light modulators (SLMs) and single-pixel detectors (SPDs) with a wide spectral response have been proposed\cite{jiang2017adaptive,jiang2018efficient,shi2019fast,zhang2019image,deng2020image,zha2021single,zha2022complementary,yang2022image,zhang2013improving,jiao2019motion,zhang2017fast,xu20181000,jiang2020imaging,hahamovich2021single,jiang2021single,monin2021single,sun2019gradual,yang2020compressive,wu2021fast,sun2022simultaneously,Yang2022Anti-motion,jiang20222,guo2022fast,xiao2022single}.

Among these methods, the image-free method can track a moving object at a very high frame rate. Zhang et al. proposed a real-time image-free tracking method based on Fourier basis patterns and achieved a temporal resolution of 1666 frames per second (fps) with a 10000-$\mathrm{Hz}$ digital micromirror device (DMD)\cite{zhang2019image}. Deng et al. extended the method to realize the three-dimensional trajectory tracking of a fast-moving object with 1666~$\mathrm{fps}$\cite{deng2020image}. Zha et al. proposed a fast-moving object tracking method based on geometric moment patterns and realized a frame rate of 7400 ~$\mathrm{Hz}$\cite{zha2021single}. Then, Zha et al. also proposed a complementary measurement scheme, which increased the frame rate of the method to 11.1~$\mathrm{kHz}$\cite{zha2022complementary}. However, the above methods cannot image a moving object while tracking the object at a high frame rate. 

Single-pixel imaging (SPI) \cite{edgar2019principles} based on SPD requires several modulation patterns to image an object, and the operating rate of the SLM for modulation is limited, resulting in a conflict between the sampling time and the reconstructed image quality. For a moving object, the sampling time allocated to a single moving frame is very short, and combining multiple moving frames for the calculation will result in motion blur. To address this problem in SPI, a moving object can be imaged by estimating the moving speed of the object using an algorithm\cite{zhang2013improving,jiao2019motion}, choosing the proper modulation patterns or increasing the speed of SLM to shorten the sampling time\cite{zhang2017fast,xu20181000,jiang2020imaging,hahamovich2021single,jiang2021single,monin2021single}, and estimating motion information based on low-resolution images\cite{sun2019gradual,yang2020compressive,wu2021fast}. In recent years, methods by estimating motion information of the moving object have been commonly used to image a moving object. Zhang et al. proposed a method for imaging a uniformly moving object by modifying patterns and velocity parameters during reconstruction\cite{zhang2013improving}. Jiao et al. proposed a method for estimating the motion parameters of a moving object under the assumption that the object motion type is known\cite{jiao2019motion}. In addition, many methods for obtaining the motion information of an object have been proposed, such as calculating cross-correlation\cite{sun2019gradual,wu2021fast} or low-order moments\cite{yang2020compressive} of images, using laterally shifting patterns\cite{sun2022simultaneously}, and projecting two-dimensional projective patterns\cite{Yang2022Anti-motion}. Even so, the frame rate of these methods is significantly lower than that of the imaging-free tracking methods. Inspired by the above methods, a concept of tracking and imaging a moving object emerges naturally: we can first determine the object's motion information using the image-free method and then transform the spatial-coding patterns of the object using motion information; when there is a sufficient number of modulating patterns, the image of the moving object can be reconstructed using the compressed sensing\cite{donoho2006compressed,candes2008introduction,duarte2008single} algorithm. Some similar ideas have been used in the most recent researches. Guo et al. combined geometric moment patterns and Hadamard patterns to achieve obtaining the relative displacements and imaging of a
moving object at a frame rate of 5.55 ~$\mathrm{kHz}$\cite{guo2022fast}. Xiao et al. achieved tracking and imaging of a fast-rotating object using Hadamard patterns and low-order geometric moment patterns\cite{xiao2022single}.

In this study, we design a new pattern sequence to achieve a high frame rate of relative displacement detection and imaging of a translational object. Four binary Fourier patterns and two differential Hadamard patterns are used to modulate one frame of the object, and then the modulated light signals are obtained by SPD. The displacement of the moving object for each moving frame can be determined by these six detection values. Based on the determined displacements and patterns, we can recalculate the reconstruction matrix and reconstruct the moving object image. The frame rate of obtaining the relative displacements of a moving object using this pattern sequence can match that of the image-free method in Ref.[\citenum{zhang2019image}]. The proposed method is verified through both simulations and experiments.

\section{Method}
In Fourier SPI (FSPI)\cite{zhang2015single}, the spatial information of an object is encoded by an SLM using Fourier basis patterns, and the series of modulated total light intensities are detected by an SPD. The required Fourier basis patterns are typically described by a pair of spatial frequencies and an initial phase. A Fourier basis pattern $P(x, y)$ can be represented by its corresponding spatial frequency pair $(f_x, f_y)$ and corresponding initial phase $\phi _0$: 
\begin{equation}
	\label{eq:fs}
	P\left(x, y \mid f_{x}, f_{y}, \varphi_{0}\right)=a_0+b_0 \cos \left[2 \pi\left(f_{x} x+f_{y} y\right)+\varphi_{0}\right],
\end{equation}
where $a_0$ represents the average intensity of the Fourier basis pattern, $b_0$ represents the contrast of the basis pattern, and $(x, y)$ corresponds to the two-dimensional spatial coordinates of the basis pattern. The modulated total light intensity $I$ can be obtained using the above Fourier basis patterns to modulate the illumination light or the detection area:
\begin{equation}
	\label{eq:fimg}
	I =\sum\nolimits_{x, y} O(x, y)P(x, y), \\
\end{equation}
where $O(x, y)$ represents the object image. Based on the linear response of the SPD to the light intensity within its effective detection range, the modulated light intensity can be replaced by the value measured by the SPD. The Fourier coefficients of the corresponding Fourier domain are obtained by these measured values. The four- and three-step phase-shifting methods are two commonly used methods for obtaining Fourier coefficients in FSPI\cite{zhang2017fast}. The four-step phase-shifting method requires four Fourier basis patterns with the same spatial frequency but different phases to obtain a Fourier coefficient. These four patterns are denoted as $P(f_x, f_y,0)$, $P(f_x, f_y,\pi/2)$, $P(f_x, f_y,\pi)$, and $P(f_x, f_y,3\pi/2)$, respectively. The corresponding single-pixel values are denoted as $I_0$, $I_{\pi/2}$, $I_{\pi}$, and $I_{3\pi/2}$, respectively. Then the corresponding Fourier coefficient is given by Eq.~(\ref{eq:4step}):
\begin{equation}
	\widetilde{O}(f_x, f_y)=(I_{0}-I_{\pi})+j(I_{\pi/2}-I_{3\pi/2}).
	\label{eq:4step}
\end{equation}
Note that the patterns $P(f_x, f_y, 0)$ is the inverse of the pattern $P (f_x, f_y, \pi)$; $P (f_x, f_y, \pi/2)$ is the inverse of the pattern $P (f_x, f_y, 3\pi/2)$. Similarly, the three-step phase-shifting method requires three Fourier basis patterns with the same spatial frequency but different initial phases. These three patterns are respectively denoted as $P(f_x, f_y,0)$, $P(f_x, f_y,2\pi/3)$, and $P(f_x, f_y,4\pi/3)$. The corresponding single-pixel values are denoted as $I_0$, $I_{2\pi/3}$, and $I_{4\pi/ 3}$, and the corresponding Fourier coefficient is given by Eq.~(\ref{eq:3step}):
\begin{equation}
	\widetilde{O}(f_x, f_y)=(2I_0-I_{2\pi/3}-I_{4\pi/3})+\sqrt{3}j(I_{2\pi/3}-I_{4\pi/3}).
	\label{eq:3step}
\end{equation}
The commonly used spatial light modulator in SPI is a DMD. These Fourier basis patterns are grayscale and cannot be directly loaded on the DMD. Binarization is typically required when these patterns are used for modulation. Fourier basis pattern generation via temporal dithering or signal dithering\cite{huang2018computational} is at the expense of temporal resolution in DMD-based FSPI. The spatial dithering strategy proposed by Zhang et al. can increase the speed of FPSI by two orders of magnitude compared with the temporal dithering method \cite{zhang2017fast}. The high temporal resolution of the dithering method is important for imaging a fast-moving object. So, the grayscale patterns used in this study can be binarized using the upsampling scheme\cite{zhang2017fast} and Floyd–Steinberg dithering method\cite{floyd1976adaptive} as in Ref.~\cite{zhang2017fast}. After binarization, the pattern $P(f_x, f_y, 0)$ plus the pattern $P (f_x, f_y, \pi)$ equals the all-one pattern; the pattern $P (f_x, f_y, \pi/2)$ plus the pattern $P (f_x, f_y, 3\pi/2)$ equals the all-one pattern, as well.

In the Fourier transform, all points in the spatial domain will contribute to each coefficient in the Fourier domain; as a result, a displacement change in the spatial domain will directly affect the Fourier coefficient. Based on the linear phase shift property of the Fourier transform, Zhang et al. proposed a method for detecting the object motion trajectory using two Fourier coefficients for each frame\cite{zhang2019image}. The specific principle is that the displacement$(\Delta x,\Delta y)$ of the object image $O(x, y)$ in the spatial domain will result in a phase shift in the Fourier domain $(-2\pi f_x\Delta x,-2\pi f_y\Delta y)$, which can be expressed as
\begin{equation}
	\label{eq:cmpxy}
	O\left(x-\Delta x, y-\Delta y\right)=F^{-1}\left\{\widetilde{O}\left(f_{x}, f_{y}\right) \exp \left[-j 2 \pi\left(f_{x} \Delta x+f_{y} \Delta y\right)\right]\right\},
\end{equation}
where $(f_x, f_y)$ denotes the spatial frequency coordinate in the Fourier domain, $\widetilde{O}(f_x, f_y)$ denotes the Fourier spectrum of the original image $O(x, y)$, and $F^{-1}$ represents the inverse Fourier transform. The relative displacement of the object can be calculated by measuring the phase shift term $\varphi=-2\pi(f_x \Delta x+f_y \Delta y)$ of each frame. Finally, the displacement $\Delta x$ and $\Delta y$ are calculated by obtaining $\widetilde{O}(f_x,0)$ and $\widetilde{O}(0, f_y)$ of each frame\cite{zhang2019image}:
\begin{equation}
	\label{eq:cmpdx}
	\begin{aligned}
		&\Delta x=-\frac{1}{2 \pi f_{x}} \cdot \arg \left\{\left[\widetilde{O}\left(f_{x}, 0\right) \cdot \overline{\widetilde{O}_{\mathrm{bg}}\left(f_{x}, 0\right)}\right]\right\}, \\
		&\Delta y=-\frac{1}{2 \pi f_{y}} \cdot \arg \left\{\left[\widetilde{O}\left(0, f_{y}\right) \cdot \overline{\widetilde{O}_{\mathrm{bg}}\left(0, f_{y}\right)}\right]\right\},
	\end{aligned}
\end{equation}
where $arg\{\}$ denotes the argument operation, $^{\overline{~}}$ denotes the complex conjugate operation, $\widetilde{O}_{\mathrm{bg}}\left(f_{x}, 0\right)$ and $\widetilde{O}_{\mathrm{bg}}\left(0, f_{y}\right)$ represent the two Fourier coefficients obtained at the initial position before the object starts moving, and $\widetilde{O}(0, f_y)$ and $\widetilde{O}(f_x,0)$ represent the two Fourier coefficients obtained at the current moving frame. Six binary Fourier basis patterns for each frame can realize real-time tracking of moving object trajectories by using the three-step phase-shifting method, as verified in Ref.[\citenum{zhang2019image}]. 

\begin{figure}[htbp]
\centering
 \includegraphics[width=0.9\textwidth]{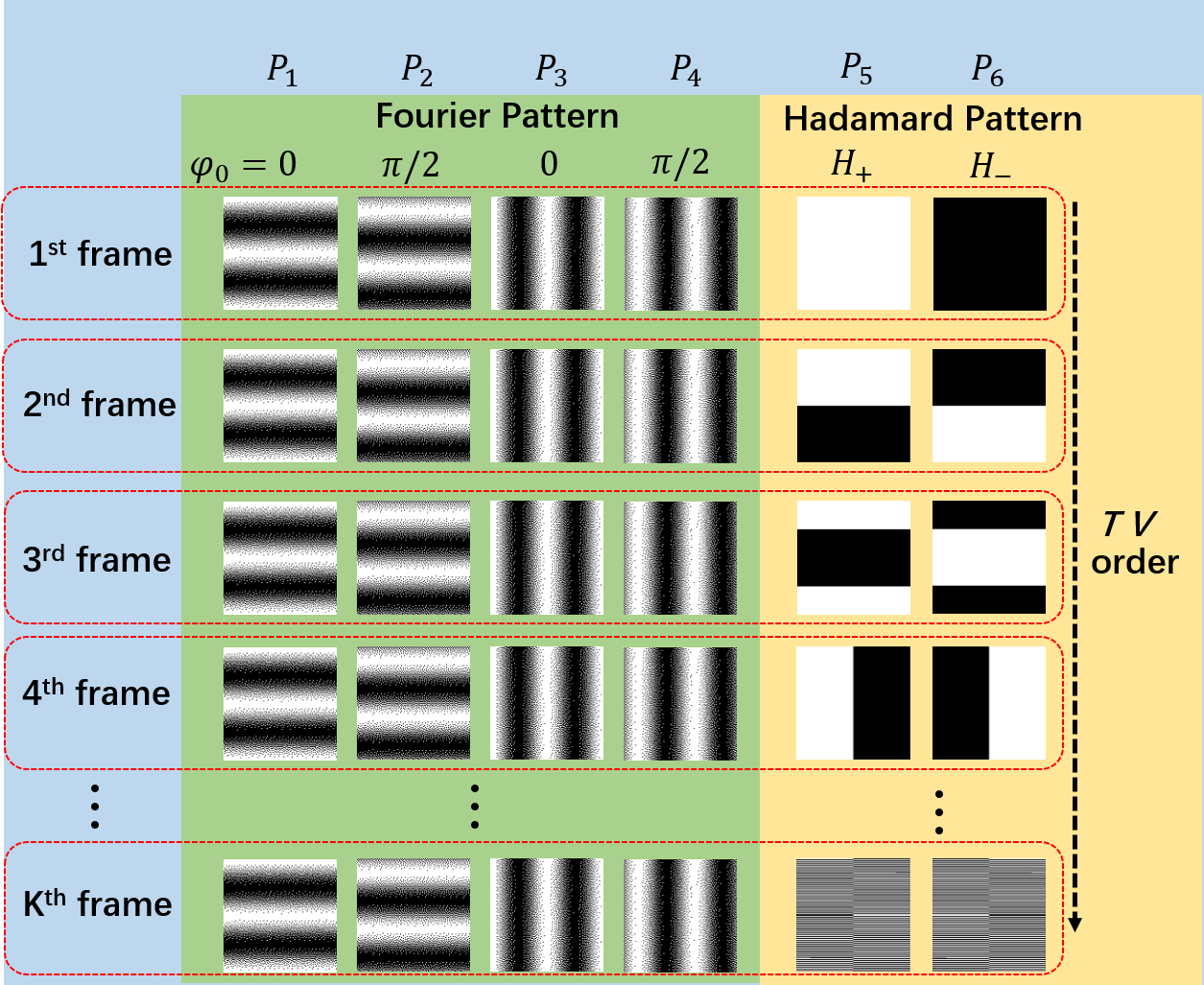}
 \caption{Schematic of pattern design. Each motion frame corresponds to six patterns, of which four are binary Fourier basis patterns and two are differential Hadamard basis patterns. The Fourier patterns of all frames are the same, and the corresponding phases are $0$ and $\pi/2$, respectively. The Hadamard patterns corresponding to different motion frames are sorted according to the total variation (TV) ordering method\cite{yu2020super}.}
\label{fgr:fig1}
\end{figure}

As shown in Fig.~\ref{fgr:fig1}, the proposed pattern sequence consists of six patterns for each frame. 
Two differential Hadamard basis patterns that encoded the object spatial information are embedded in every four Fourier basis patterns to achieve the high rate of Ref.[\citenum{zhang2019image}], and finally, the moving object is imaged. The first four binary Fourier basis patterns of all motion frames are the same which correspond to the Fourier basis patterns of $P(f_x,0,0)$, $P(f_x,0,\pi/2)$, $P(0, f_y,0)$, and $P(0, f_y,\pi/2)$, respectively. The spatial frequencies $f_x$ and $f_y$ in this study are both $2/m$, where $m \times m$ represents the spatial resolution of the image. The two differential Hadamard patterns $H_+^k$ and $H_-^k$ are calculated from the $k^{th}$ Hadamard pattern $H^k$:
\begin{equation}
\begin{aligned}
	&H_+^k=(H^k+1)/2, \\
	&H_-^k=(1-H^k)/2, 
\end{aligned}
\end{equation}
$H_+^k$ plus $H_-^k$ also equals the all-one pattern. The Hadamard patterns in each motion frame differ. Each Hadamard pattern is selected according to the total variation (TV) sorted method\cite{yu2020super}. For six patterns of each frame, the corresponding single-pixel values are $I_{x,0}$, $I_{x,\pi/2}$, $I_{y,0}$, $I_{y,\pi/2}$, $I_{H+}$, and $I_{H-}$, respectively. Based on the four-step phase-shifting method, the two Fourier coefficients are calculated by:
\begin{equation}
\begin{aligned}
	&\widetilde{O}(f_x, 0)=(2I_{x,0}-I_{H+}-I_{H-})+j(2I_{x,\pi/2}-I_{H+}-I_{H-}), \\
	&\widetilde{O}(0, f_y)=(2I_{y,0}-I_{H+}-I_{H-})+j(2I_{y,\pi/2}-I_{H+}-I_{H-}),
\end{aligned}
\end{equation}
and the Hadamard coefficient can be calculated by:
\begin{equation}
		I_H=I_{H+}-I_{H-}.
\end{equation}
\begin{figure}[htbp]
\centering
 \includegraphics[width=0.5\textwidth]{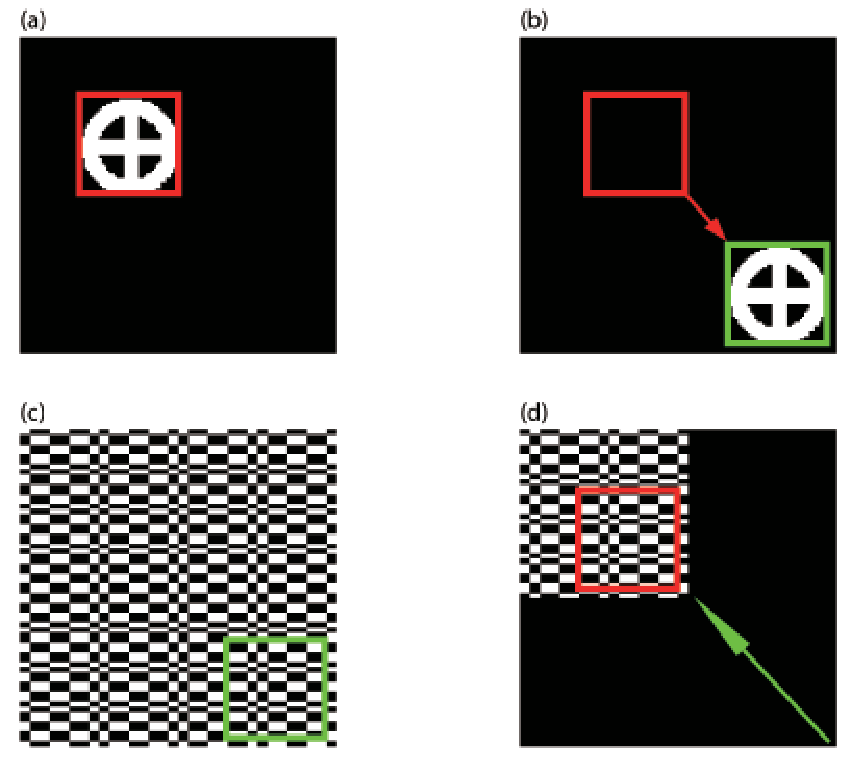}
 \caption{Schematic of modulating pattern transformation. (a) the initial position of the object; (b) the position of the object after movement; (c) the modulating pattern of the object after movement; (d) the transformed modulation pattern.}
\label{fgr:fig2}
\end{figure}
The displacement of the object can be calculated by Eq.~(\ref{eq:cmpdx}), and then the displacement of the current frame can be determined. In addition, four binary Fourier basis patterns, one Hadamard basis pattern, and the corresponding five single-pixel values are obtained for each frame, which can be used in the final imaging procedure. We normalize the measured values by the total light intensity of each frame to mitigate the influence of non-uniform illumination during the object movement. For the six patterns of each frame, the corresponding single-pixel detection values are marked as $I_i (i=1,2,3,4,5,6)$. $I_5$ and $I_6$ are the intensities of two differential Hadamard patterns, and the sum of them is the total intensity of the frame. The normalized value $\Tilde{I_i}$ can be expressed as:
\begin{equation}
		\Tilde{I_i}=\frac{I_i}{I_5+I_6},i=1,2,3,4,5,6.
\end{equation}
The displacement of the object during the pattern modulation is equivalent to that the object being static and the pattern moving in the opposite direction to the object's motion. The object image can be reconstructed from the recorded single-pixel values and transformed patterns when transformed patterns are sufficient. The process of transforming patterns is shown in Fig.~\ref{fgr:fig2}. The whole pattern of the moved object requires reverse translation according to the calculated relative displacement. The total variation augmented lagrangian alternative direction algorithm (TVAL3)\cite{li2013efficient} is an efficient and widely used compressed sensing\cite{donoho2006compressed,candes2008introduction,duarte2008single} algorithm. The TVAL3 solver can be employed to reconstruct the object using the transformed pattern sequence and corresponding single-pixel values.

\section{Results and discussion} 
\subsection{Simulations}
\begin{figure}[htbp]
\centering
 \includegraphics[width=0.9\textwidth]{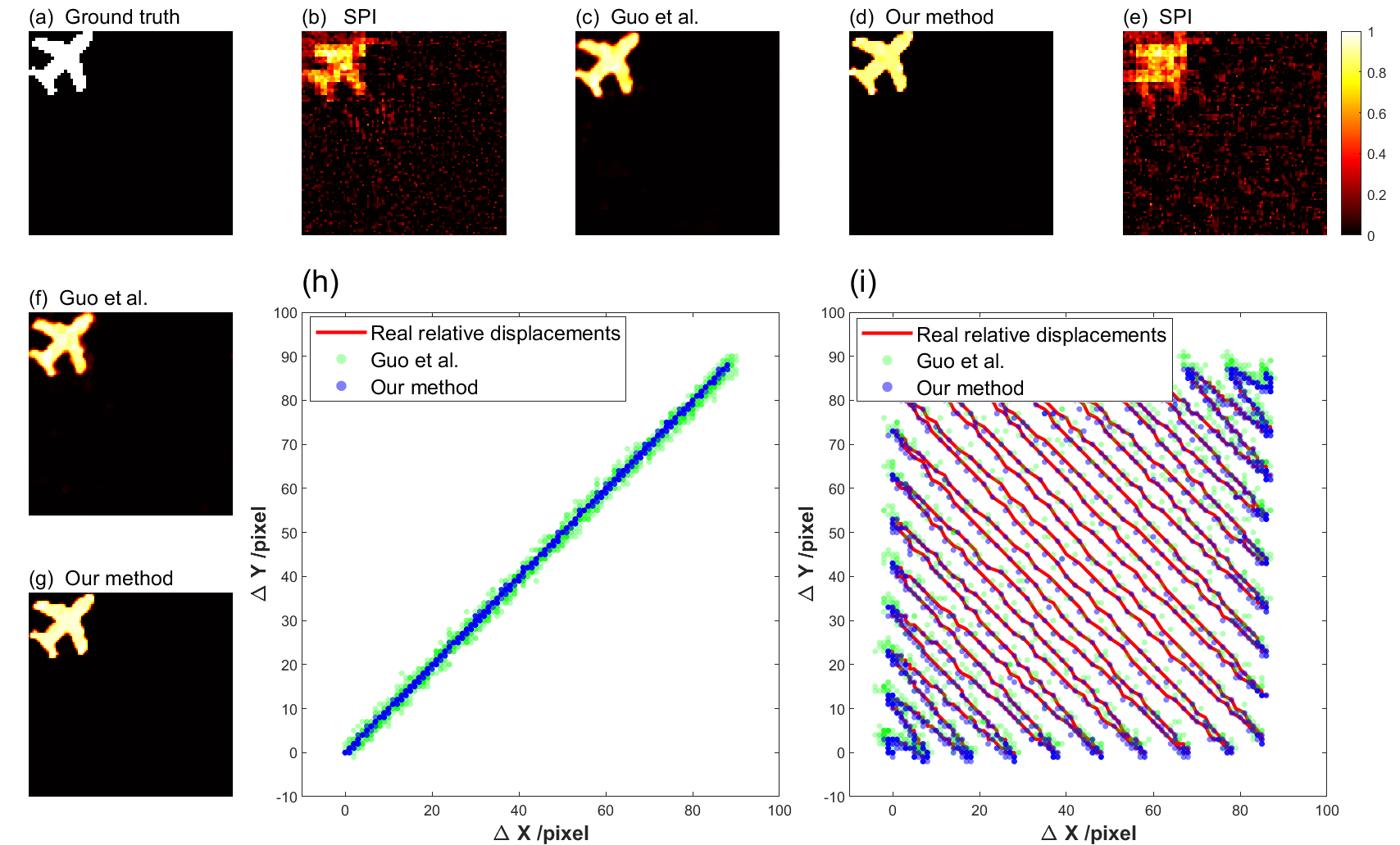}
 \caption{Simulation results of one object with two trajectories. (a) Simulated object. (d) and (e) show the reconstructed images by the conventional SPI method. Full comparisons under different noise levels are shown in Supplement 1. (c) and (f) show the reconstructed images by Guo et al.'s method\cite{guo2022fast}. (d) and (g) show the reconstructed images by the proposed method. (h) Relative displacements of the type-\uppercase\expandafter{\romannumeral1} trajectories reconstructed by the two methods; (i) Relative displacements of the type-\uppercase\expandafter{\romannumeral2} trajectories reconstructed by the two methods.}
\label{fgr:fig3}
\end{figure}

In the simulation, an object with two trajectories is simulated to verify the proposed method. The object image is shown in Fig.~\ref{fgr:fig3}(a), and the resolution is $128 \times 128$ pixels. The total number of moving frames is 1666, and each frame corresponds to six patterns, meaning that a total of 9996 patterns are used in the simulation. Guo et al.'s method based on geometric moment patterns\cite{guo2022fast} is also simulated for comparison. The geometric moment patterns used in the simulation were binarized using the Floyd–Steinberg dithering algorithm\cite{floyd1976adaptive} with an upsampling ratio of 2. We studied the relationship between the upsampling ratio of spatial dithering and the relative position accuracy of our method. It was found that our method is not sensitive to the upsampling ratio, so we chose an upsampling ratio of 1 without sacrificing the spatial resolution of the image (see Supplement 1). The total number of patterns for Guo et al.'s method\cite{guo2022fast} is 6664 corresponding to 1666 moving frames. Note that Guo et al.'s method\cite{guo2022fast} can obtain the centroids of the object in each moving frame, so the displacements of the object can be calculated by subtracting the centroid coordinates of the first frame. Hadamard patterns are chosen to modulate and reconstruct the image of the moving object as in the conventional SPI method, and 9996 differential Hadamard patterns are used according to the TV order\cite{yu2020super} for improved quality. Gaussian white noise with $\sigma=0.1$ is included in the measurement for simulating real noisy experiments.

The mean square error (MSE) is introduced to evaluate the accuracy of the reconstructed relative displacement. The MSE of the reconstructed relative displacement coordinate $Y$ and the original coordinate $X$ is defined as follows:
\begin{equation}
	MSE(X, Y)=\frac{1}{n}\sum_{i=1}^{n}(X_i-Y_i)^2,
\end{equation}
where $n$ represents the total number of frames. The smaller the MSE, the closer the reconstructed relative displacement is to the original relative displacement. The peak signal-to-noise ratio (PSNR) and structural similarity (SSIM) \cite{wang2004Iamge} are introduced to evaluate the quality of reconstructed images. The PSNR between the original image $x$ and the reconstructed image $y$ is defined as follows:
\begin{equation}
	PSNR\left(x, y\right)=10\log \frac{peakval^{2}}{MSE\left(x, y\right)},
	\label{eqn:psnr}
\end{equation}
where $MSE (x, y)$ represents the MSE between $x$ and $y$. $Peakval$ is the maximum value of the image data type. The larger the PSNR value, the higher the reconstruction quality. The SSIM between the original image $x$ and the reconstructed image $y$ is defined as follows:
\begin{equation}
  SSIM\left(x,y\right)=\frac{\left(2\mu_x \mu_y+C_1\right)\left(2\sigma_{xy}+C_2\right)}{\left(\mu^2_x +\mu^2_y+C_1\right)\left(\sigma^2_x+\sigma^2_y+C_2\right)},
  \label{eqn:ssim}
\end{equation}
where $\mu_x$ represents the mean of $x$, $\sigma_x$ represents the variance of $x$, $\sigma_{xy}$ represents the covariance of $x$ and $y$, $\mu_y$ represents the mean of $y$, $\sigma_y$ represents the variance of $y$, $C_1$ and $C_2$ are constants. The value range of SSIM is $[0, 1]$. The larger the SSIM value, the higher the structural similarity between the two images.

The simulations with different noise distributions were repeated five times. Figure~\ref{fgr:fig3} illustrates the results of the corresponding method. Figure~\ref{fgr:fig3}(h) and Figure~\ref{fgr:fig3}(i) compare the displacement relative displacements reconstructed by the two methods and the original relative displacements.
For trajectory (type-\uppercase\expandafter{\romannumeral1}) in Fig.~\ref{fgr:fig3}(h), the image reconstructed by conventional SPI, Guo et al.'s method\cite{guo2022fast}, and the proposed method are depicted in Fig.~\ref{fgr:fig3}(b), Fig.~\ref{fgr:fig3}(c), and Fig.~\ref{fgr:fig3}(d), respectively. For trajectory (type-\uppercase\expandafter{\romannumeral2}) in Fig.~\ref{fgr:fig3}(i), the images reconstructed by conventional SPI, Guo et al.'s method\cite{guo2022fast}, and the proposed method are depicted in Fig.~\ref{fgr:fig3}(e), Fig.~\ref{fgr:fig3}(f), and Fig.~\ref{fgr:fig3}(g), respectively. Table~\ref{tbl:noise_sim} shows the mean MSEs, PSNRs, and SSIMs of these methods. 
\begin{table}[htb]
	\small
	\centering
	\caption{Comparisons of reconstructed relative displacements and images using different methods }
	\label{tbl:noise_sim}
	\begin{tabular*}{0.9\textwidth}{@{\extracolsep{\fill}}ccccc}
		\hline
		\multirow{2}*{Trajectory}& \multirow{2}*{Method}&{Restored relative displacements}&\multicolumn{2}{c}{Reconstructed images} \\ \cline{3-5} 
		& & {MSE} & {PSNR} & {SSIM} \\ \hline
		\multirow{4}*{Type-\uppercase\expandafter{\romannumeral1}} &{Conventional SPI} &{N/A} &{16.5194} &{0.1345} \\ 
		& {Guo et al.\cite{guo2022fast}} & {4.2483} & {20.3526} &{0.9085} \\
		& {Our method} & {\textbf{0.5155}}  &{\textbf{24.6712}} & \textbf{0.9735}\\ \hline
		\multirow{4}*{Type-\uppercase\expandafter{\romannumeral2}} &{Conventional SPI} & {N/A} &{16.1325}& {0.1248} \\ 
		& {Guo et al.\cite{guo2022fast}} & {4.0756}& {20.7173} &{0.9132}\\
		& {Our method} & {\textbf{0.3102}}  &{\textbf{26.1158}} & {\textbf{0.9805}} \\ \hline
	\end{tabular*}
\end{table}
For a more detailed comparison under different noise levels, see Supplement 1. From the above results, the proposed method can effectively reconstruct the relative displacements of the moving object in the presence of noise. Guo et al.'s method \cite{guo2022fast} exhibits a significant deviation in the reconstructed relative displacements, indicating that the geometric moment patterns are more sensitive to noise than the Fourier patterns. The reconstructed images of the conventional SPI method are blurred and degraded because of the motion of the object during the measurement process, whereas the proposed method can effectively reconstruct the images of the moving object with higher quality than Guo et al.'s method\cite{guo2022fast}.

\begin{figure}[htbp]
\centering
 \includegraphics[width=0.8\textwidth]{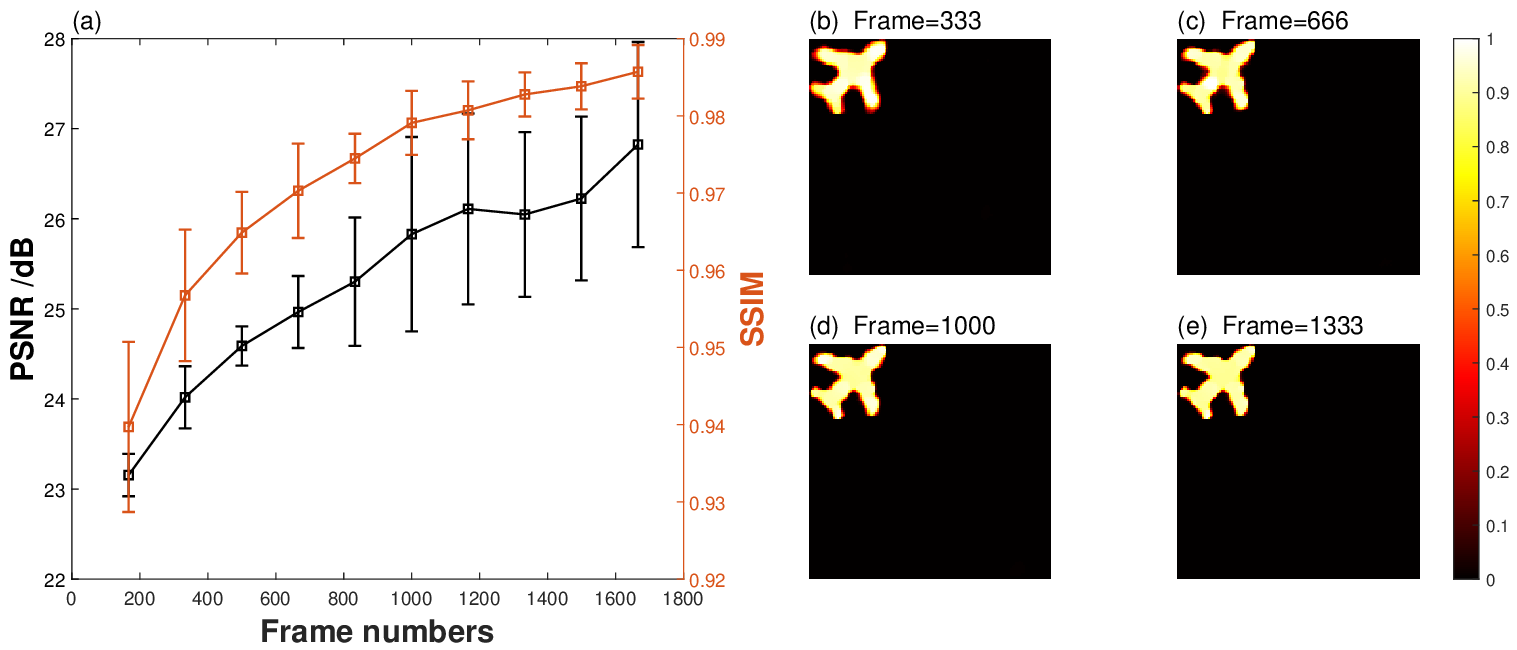}
 \caption{Influence of the number of moving frames on the imaging quality computed using the proposed method. (a) PSNRs and SSIMs of the reconstructed images using different numbers of frames ; (b–e) reconstructed images using 333, 666, 1000, and 1333 frames, respectively.}
\label{fgr:fig4}
\end{figure}
The influence of the number of moving frames on the imaging quality should also be considered by reconstructing the image while sequentially reducing the number of frames. The reconstructed images obtained while moving along the type-\uppercase\expandafter{\romannumeral2} trajectory were investigated in this simulation. Ten groups of data were selected, including 10\%, 20\%, ..., 90\%, and 100\% of the total 1666 frames, respectively, to compute the reconstructed images. The corresponding PSNRs and SSIMs are shown in Fig.~\ref{fgr:fig4}(a). Figures \ref{fgr:fig4}(b–e) depict the reconstructed images with the corresponding frame numbers 333, 666, 1000, and 1333. The results indicate that, considering the influence of noise, when the number of sampling frames is greater than a certain frame number (e.g., 666 frames), the moving object image with good quality can be reconstructed by our method. 

\subsection{Experiments}
\begin{figure}[htbp]
\centering
 \includegraphics[width=1\textwidth]{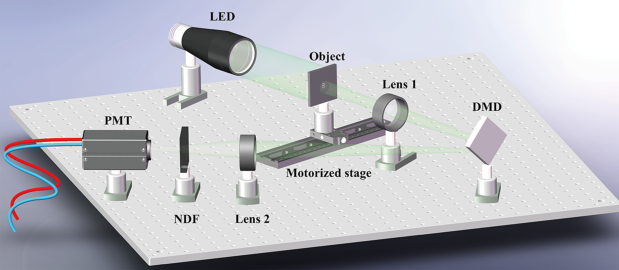}
 \caption{Experimental setup. A motorized stage moves the object in one direction. The transmitted object is imaged on the digital micromirror device (DMD) after being illuminated by the light-emitting diode (LED) source. After being modulated by the DMD, the modulated light is collected into the photomultiplier tube (PMT) and converted into measurement values through the data acquisition board. NDF: neutral density filter.}
\label{fgr:fig5}
\end{figure}

The proposed method is verified through experiments. The experimental system device consists of a light-emitting diode (LED) source with a maximum power of 5~$\mathrm{W}$, a linear motorized stage (KA400Z, 
Zolix), a DMD (Texas Instruments DLP7000), a photomultiplier tube (PMT, H10682-210, Hamamatsu Photonics), and a data acquisition board, as depicted in Fig.~\ref{fgr:fig5}. The modulation patterns are preloaded into the DMD in advance for modulation. The transmitted object is imaged on the DMD after being illuminated by the light source. After being modulated by the DMD, the modulated light is collected into the PMT and converted into measurement values through the data acquisition board. The DMD operated at a high refresh rate of 20000~$\mathrm{Hz}$. The pixel size of the modulation patterns located on the DMD was $256 \times 256$ pixels, and all $2 \times 2$ pixels were merged into one super pixel, meaning that the image size of the moving object was $128 \times 128$ pixels. The Fourier basis patterns used in the method were binarized using the Floyd–Steinberg dithering algorithm\cite{floyd1976adaptive} with an upsampling ratio of 1. In each experiment group, 9996 patterns were used corresponding to frame numbers of 1666, and the total measurement time was 0.4998~$\mathrm{s}$ for a modulating rate of 20000~$\mathrm{Hz}$. Guo et al.'s method\cite{guo2022fast} was also applied for comparison. The geometric moment patterns used in the experiments were binarized using the Floyd–Steinberg dithering algorithm\cite{floyd1976adaptive} with an upsampling ratio of 2. The total number of patterns for 
Guo et al.'s method\cite{guo2022fast} was 6664 corresponding to 1666 moving frames and the total measurement time was also 0.4998~$\mathrm{s}$ for a modulating rate of 13333~$\mathrm{Hz}$.

\subsubsection{Background-free situation}
\begin{figure}[htbp]
\centering
 \includegraphics[width=0.9\textwidth]{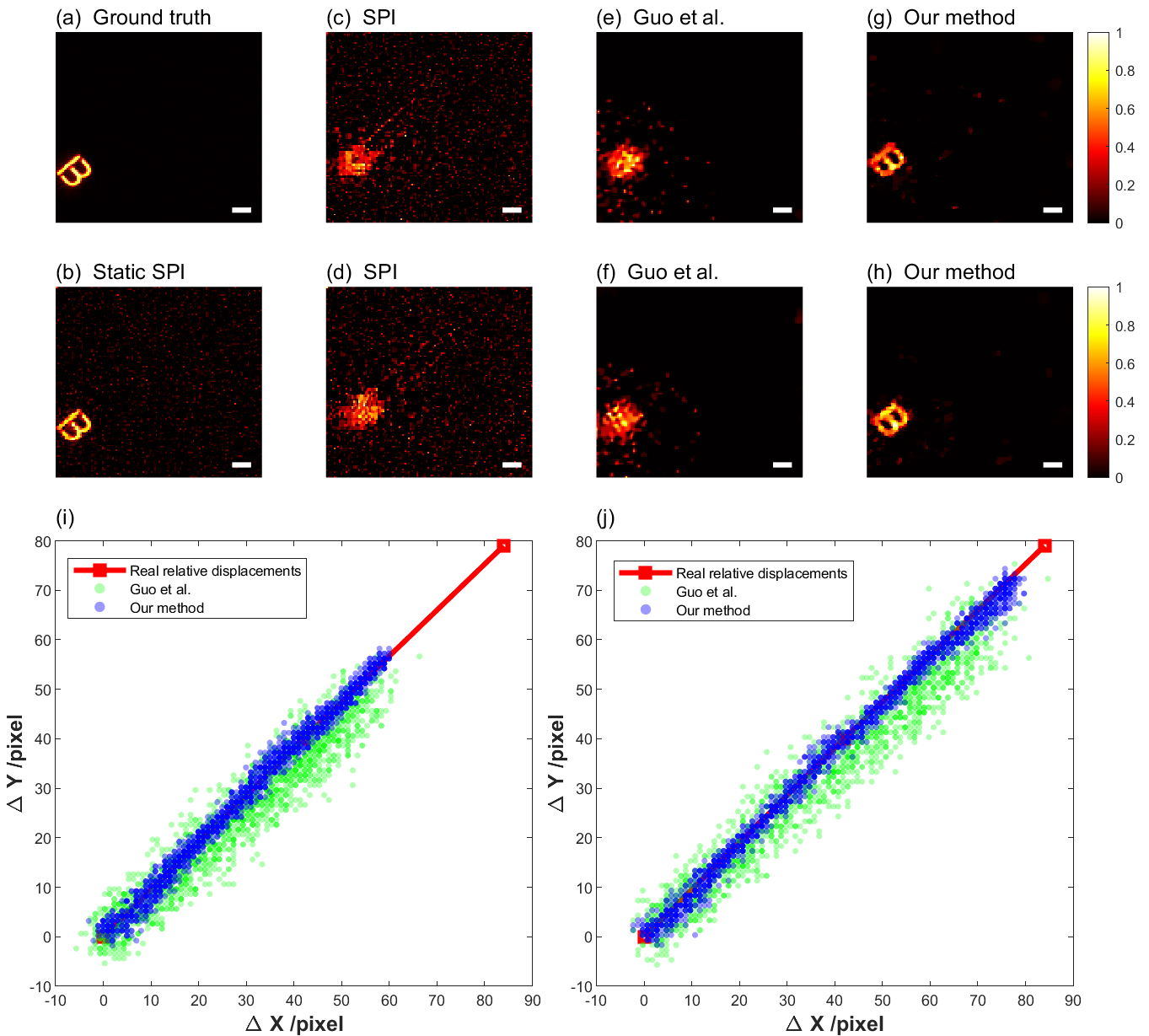}
 \caption{Imaging and obtaining the relative displacements results of an object with two moving speeds. (a) shows the Ground truth image. (c) and (d) show the reconstructed moving objects by the conventional SPI method. (e) and (f) show the reconstructed moving objects by Guo et al.'s method\cite{guo2022fast}. (g) and (h) show the reconstructed moving objects by the proposed method. (i) and (j) compare the real relative displacements and calculated relative displacements using two methods. Scale bar: 2~$\mathrm{mm}$.}
\label{fgr:fig6}
\end{figure}
In this section, the moving object was a transmitted object "B" with a size of $2.5~\mathrm{mm} \times 3.2 ~\mathrm{mm}$. In the first group experiment, the object moved straightly along the diagonal of the image with a constant speed of 25~$\mathrm{mm/s}$. In the second group experiment, the object moved straightly along the diagonal of the image with an initial speed of 25~$\mathrm{mm/s}$ and an accelerated speed of 50~$\mathrm{mm/s^2}$. The Ground truth image was obtained by imaging the static object using the conventional SPI method, as depicted in Fig.~\ref{fgr:fig6}(a). To obtain a clear static image, the DMD was operated at a modulating rate of 100~$\mathrm{Hz}$ and a total of 19992 differential Hadamard patterns were used corresponding to a sample ratio of 61.01\%. The static object image at a modulating rate of 20000~$\mathrm{Hz}$, as depicted in Fig.~\ref{fgr:fig6}(b), illustrates the image degradation under high frame rate sampling. The real relative displacement of the object was determined by imaging the static object along the displacement axis of the motorized stage. Three static object images reconstructed using the conventional SPI method were combined to obtain the real relative displacement. The reconstructed images for moving object using the conventional SPI method are illustrated in Fig.~\ref{fgr:fig6}(c) and Fig.~\ref{fgr:fig6}(d), which are blurred due to motion. The reconstructed images for moving object using Guo et al.'s method\cite{guo2022fast} are illustrated in Fig.~\ref{fgr:fig6}(e) and Fig.~\ref{fgr:fig6}(f), which are also blurred due to the larger deviation in the reconstructed relative displacements. Two images of the moving object with good quality were calculated using the proposed method, as shown in Fig.~\ref{fgr:fig6}(g) and Fig.~\ref{fgr:fig6}(h). Figure~\ref{fgr:fig6}(i) and Figure~\ref{fgr:fig6}(j) compare the real and calculated relative displacements using Guo et al.'s method\cite{guo2022fast} and the proposed method. The reconstructed relative displacements using the proposed method are closer to the real relative displacements, indicating that our method is more robust to noise than Guo et al.'s method\cite{guo2022fast}. Compared with the compressed sensing algorithm, which requires a long calculation time, we used the non-iterative differential ghost imaging (DGI)~\cite{ferri2010} algorithm to reconstruct the object image in Supplement 1. Although there is large background noise in the images, our method can distinguish the object, while the other two methods cannot.

\begin{figure}[htbp]
\centering
 \includegraphics[width=0.8\textwidth]{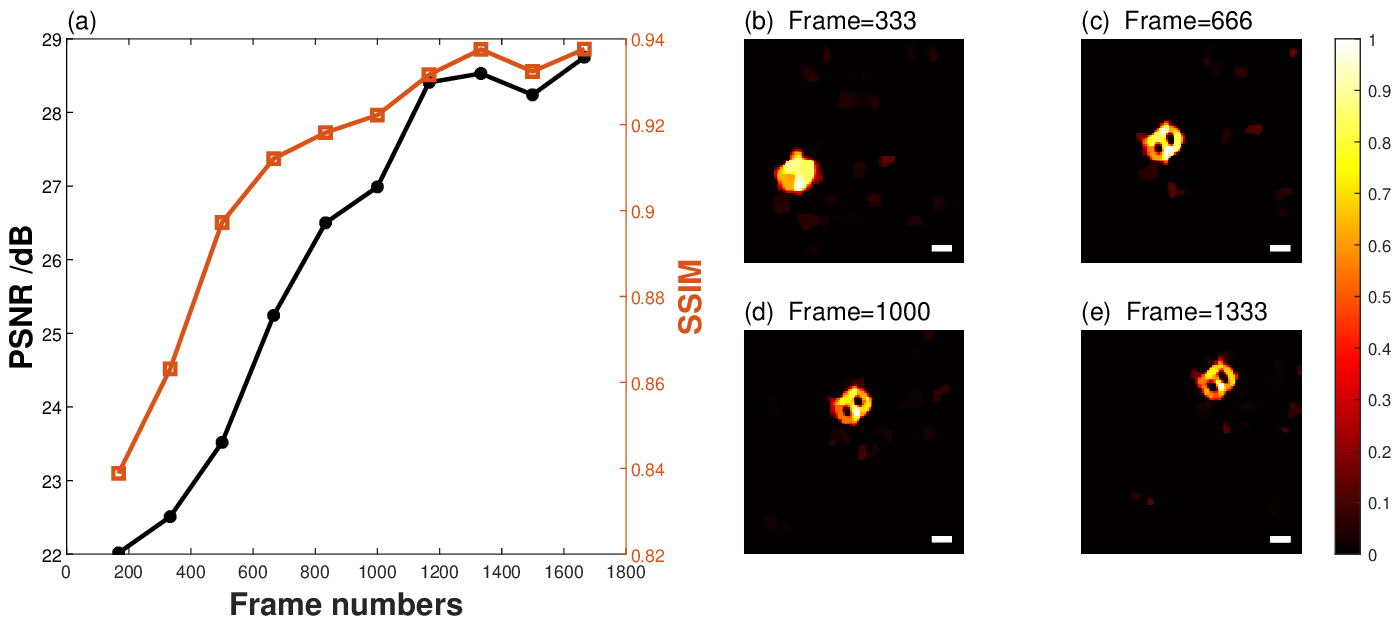}
 \caption{Experiment with the contribution of the number of measured moving frames to reconstructed imaging quality (see Visualization 1). (a) PSNRs and SSIMs of reconstructed images using different numbers of frames; (b–e) Reconstructed images using 333, 666, 1000, and 1333 frames, respectively. Scale bar: 2~$\mathrm{mm}$.}
\label{fgr:fig7}
\end{figure}
The influence of the number of moving frames on the imaging quality was investigated in this experiment. Ten groups of measured data were selected, including 10\%, 20\%, ..., 90\%, and 100\% of the total 1666 frames, respectively, to compute the reconstructed images. The corresponding PSNRs and SSIMs are depicted in Fig.~\ref{fgr:fig7}(a). The results are similar to the simulation results. When the number of sampling frames is greater than a certain frame number, our method can reconstruct the moving object image with high quality. The reconstructed images with the corresponding frame numbers 333, 666, 1000, and 1333 (corresponding to a measurement time of 0.0999~$\mathrm{s}$, 0.1998~$\mathrm{s}$, 0.3000~$\mathrm{s}$, and 0.3999~$\mathrm{s}$, respectively) are depicted in Fig.~\ref{fgr:fig7}(b–e). Visualization 1 demonstrates the motion of the object and the reconstructed image using different numbers of frames. For studying the influence of the frame rate of the relative positions obtained on the image quality, we compared the reconstructed image quality at different moving speeds. The image quality of our method is better than Guo et al.'s \cite{guo2022fast} method in all cases, demonstrating the superiority of our method. Details are described in Supplement 1.

\subsubsection{Moving object with stationary background}
\begin{figure}[htbp]
\centering
 \includegraphics[width=0.9\textwidth]{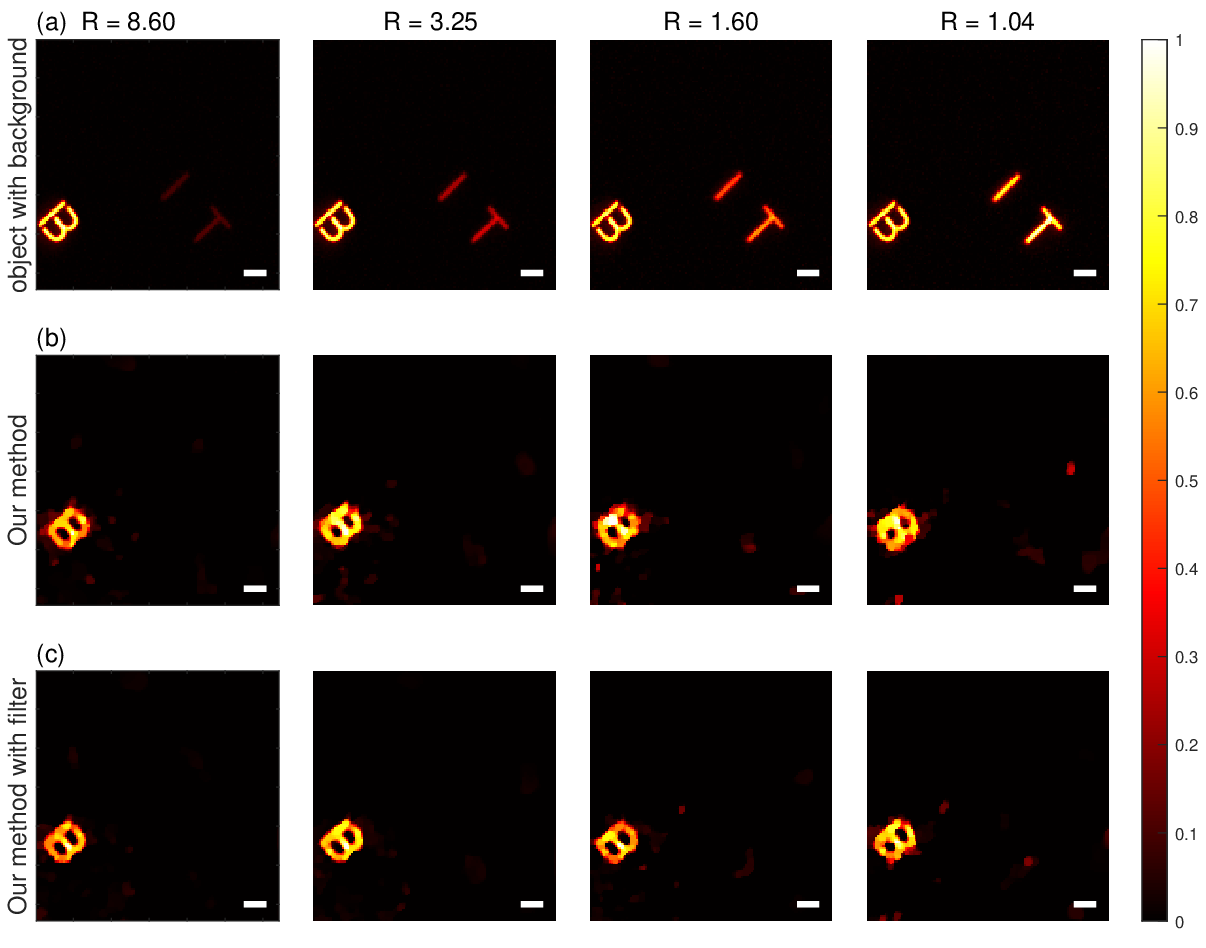}
 \caption{Experimental images of a moving object under different background intensities. (a) Images of the static object under different background intensities. (b) Reconstructed moving objects by our method. (c) Reconstructed moving objects by our method with a filter. Scale bar: 2~$\mathrm{mm}$.}
\label{fgr:fig8}
\end{figure}
\begin{figure}[htbp]
\centering
 \includegraphics[width=1\textwidth]{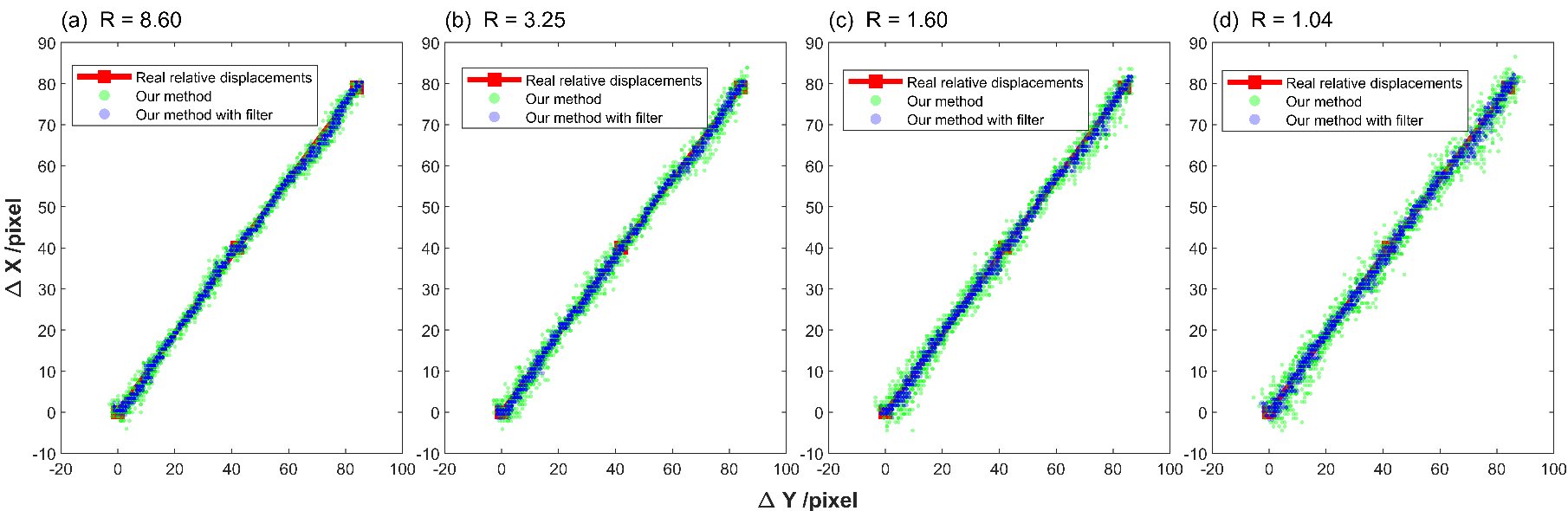}
 \caption{Calculated relative displacements of a moving object under different background intensities using our method. (a-d) compare the real relative displacements and calculated relative displacements with or without a filter under the object-to-background ratio of 8.60, 3.25, 1.60, and 1.04, respectively.}
\label{fgr:fig9}
\end{figure}
In this experiment, the moving target was also the transmitted object "B", while the stationary background consisted of a transmitted letter "I" and "T". Light could be transmitted through both object "B" and background "I" and "T" when the object move straightly along the diagonal of the image with an initial speed of 25~$\mathrm{mm/s}$ and an accelerated speed of 50~$\mathrm{mm/s^2}$. When the object began to move, the detection module began counting until the object completely moves out of the field of view. We calculated the relative displacement and image of the moving object by subtracting the measured data from the background. The ratio between the total intensity of the object and the total intensity of the background was marked as R. We performed four experiments under different background intensities ( R= 8.60, 3.25, 1.60, and 1.04, respectively) to evaluate the influence of background intensity on reconstructed relative displacement and image. Four clear images of the static object with different background intensities were obtained using a modulating rate of 100~$\mathrm{Hz}$, as depicted in Fig.~\ref{fgr:fig8}(a). The reconstructed images and relative displacements using our method are shown in Fig.~\ref{fgr:fig8}(b) and Fig.~\ref{fgr:fig9}, respectively. The high-quality images are reconstructed under a weak background. With the increase of background intensity, the quality of reconstructed images and relative displacements using our method decreases gradually. To mitigate the impact of noise, we apply a five-order mean filter to the calculated Fourier coefficients during reconstruction. The reconstructed images and relative displacements using filtered data, as shown in Fig.~\ref{fgr:fig8}(c) and Fig.~\ref{fgr:fig9}, respectively, demonstrate the improvement of the relative displacements' accuracy and imaging quality. 

\subsection{Discussion}
The key feature of the proposed method is to obtain object motion information determined using the image-free tracking method. The higher the frame rate and the more accurate the calculated relative displacement, the more accurate the reconstructed image of the moving object. In our settings, six modulation patterns were used for each motion frame. Thus, the maximum frame rate of our method is the refresh rate of DMD divided by six. Using a DMD with a refresh rate of 20000~$\mathrm{Hz}$, we achieved obtaining the relative displacements at a frame rate of 3332~$\mathrm{Hz}$ and finally imaged the moving object in the experiments. A higher frame rate can be achieved by using an SLM with a higher refresh rate. Although a higher frame rate of obtaining the relative displacements can also be achieved using geometric moment patterns, the simulation indicates that binary geometric moment patterns\cite{guo2022fast} lack differential measurements and require a higher upsampling ratio when binarizing, resulting in the loss of spatial resolution and poor robustness to noise. The four-step phase-shifting method typically exhibits better noise resistance than the three-step phase-shifting method to obtain a Fourier coefficient in FSPI\cite{zhang2017hadamard}. To have a better trade-off between the relative displacement accuracy and image quality, six measured values are used to calculate the Fourier coefficients in this study rather than the eight values required by the four-step phase-shifting method. Nevertheless, our method can reconstruct the image of a translational object with high quality without sacrificing spatial resolution compared with the method based on geometric moment patterns. More binary Fourier patterns can be added to obtain a more accurate relative displacement; however, this will decrease the maximum frame rate of obtaining the relative displacements. To mitigate the impact of noise, a mean filter can be used to reduce the fluctuation of the reconstructed relative displacement and obtain high-quality images. The influence of the number of moving frames on imaging quality is also considered in this study. The simulated and experimental results indicate that with an increase in the number of measurement frames, the quality of the reconstructed image will increase rapidly and then tends to be stable. The result also indicates that the total detection time of 0.4998~$\mathrm{s}$ is unnecessary for imaging such an object.

We also acknowledge that the proposed method has several limitations. First, it can only image a transnational object with a simple background in the field of view due to the characteristics of the Fourier transform. When the object is rotating or deforming, our method would be invalid. The method based on geometric moment patterns can obtain the rotation state of objects by adding low-order geometric moment patterns, such as the method in Ref.~\cite{xiao2022single}, and achieve tracking and imaging of a rotating object. Second, our method only can obtain the relative positions of the translational object. Without prior knowledge of the initial position of the moving object, we will not obtain the trajectory information. The limitation of our method can be improved by introducing a small number of geometric moment patterns to determine the centroid coordinates at the beginning of the tracking process so that the tracking ability can be approximately obtained through the measured centroid coordinates and relative displacements. Third, our method cannot reconstruct the image if the time of the object staying in the field of view is too short to obtain sufficient useful measurements. Compared with our method, the method based on geometric moment patterns can achieve a higher position frame rate. When the object moves at a high speed, this method of obtaining the object’s position at a high frame rate may have advantages. Considering its poor noise robustness, it may get good results when combined with our method. In addition, the image cannot be reconstructed in real time because of the use of the iterative algorithm. The non-iterative algorithm DGI is also used in Supplement 1, but there is large background noise in the restored image. In future work, the restored algorithms based on deep learning\cite{higham2018deep,Gibson2020single} may be a good choice.

\section{Conclusion}

A single-pixel detection method for restoring the relative displacements and imaging a translational object is proposed in this study. The displacement of a moving object for each moving frame can be determined by four Fourier patterns along with two differential Hadamard patterns. Based on the determined displacements and patterns, we can recalculate the reconstruction matrix and reconstruct the image of the moving object. This method does not require any prior knowledge of the object and its motion. It has been verified by simulations and experiments, achieving a frame rate of 3332 ~$\mathrm{Hz}$ to acquire relative displacements of a translational object at a spatial resolution of 128 × 128 pixels by using a 20000-$\mathrm{Hz}$ DMD. Future studies can focus on improving the frame rate for acquiring relative displacements and accelerating the reconstruction process to finally realize real-time imaging.

\begin{backmatter}
\bmsection{Funding}
Beĳing Institute of Technology Research Fund Program for Young Scholars (Grant no.20212012). 
\bmsection{Disclosures}
The authors declare no conflicts of interest.

\bmsection{Data availability} Data underlying the results presented in this paper are not publicly available at this time but may be obtained from the authors upon reasonable request.

\end{backmatter}

%%%%%%%%%% If using BibTeX:
\bibliography{sample}

\end{document}